\documentclass{article}
\usepackage[letterpaper, portrait, margin=1.5in]{geometry}
\usepackage{url}  
\usepackage{graphicx}  

\usepackage{amsmath}
\usepackage{algorithm}
\usepackage{algpseudocode}
\usepackage{url}
\usepackage{subfig}

\algdef{SE}[DOWHILE]{Do}{DoWhile}{\algorithmicdo}[1]{\algorithmicwhile\ #1 }
\DeclareMathOperator*{\argmin}{argmin}

\begin{document}
	%
	\title{Scaling Author Name Disambiguation with CNF Blocking}
	\date{}
\author{
	\medskip
	Kunho Kim$^{\dag}$, Athar Sefid$^{\dag}$, C. Lee Giles$^{\dag\ddag}$\\
	\medskip
	\scalebox{0.8}{
		\begin{tabular}{c}
			$^{\dag}$Computer Science and Engineering \\
			$^{\ddag}$Information Sciences and Technology \\ 
			The Pennsylvania State University \\
			University Park, PA 16802, USA \\
	\end{tabular}} \\
	\scalebox{0.8}{
		\begin{tabular}{c}
			kunho@cse.psu.edu, azs5955@psu.edu, giles@ist.psu.edu
	\end{tabular}}
}
	\maketitle
	\begin{abstract}
		An author name disambiguation (AND) algorithm identifies a unique author entity record from all similar or same publication records in scholarly or similar databases. Typically, a clustering method is used that requires calculation of similarities between each possible record pair. However, the total number of pairs grows quadratically with the size of the author database making such clustering difficult for millions of records. One remedy for this is a blocking function that reduces the number of pairwise similarity calculations. Here, we introduce a new way of learning blocking schemes by using a conjunctive normal form (CNF) in contrast to the disjunctive normal form (DNF). We demonstrate on PubMed author records that CNF blocking reduces more pairs while preserving high pairs completeness compared to the previous methods that use a DNF with the computation time significantly reduced. Thus, these concepts in scholarly data can be better represented with CNFs. Moreover, we also show how to ensure that the method produces disjoint blocks so that the rest of the AND algorithm can be easily paralleled. Our CNF blocking tested on the entire PubMed database of 80 million author mentions efficiently removes 82.17\% of all author record pairs in 10 minutes.
	\end{abstract}
	
	\section{Introduction}
	The author name disambiguation (AND) is a problem of identifying each unique author entity record from all publication records in scholarly databases \cite{ferreira2012brief}. It can be thought as a special case of named entity recognition \cite{nadeau2007survey} and name entity linking \cite{hirschman1998muc}, which recognizes same entities from structured data rather than free text. It is an important pre-processing step for a variety of problems. One example is processing author-related queries properly (e.g., identify all of a particular author's publications) in a digital library search engine. Another is to calculate author-related statistics, such as h-index and studying collaboration relationships between authors.
	
	Typically a clustering method is used to process the AND. Clustering requires calculating pairwise similarities between each possible pair of records, to determine whether each pair should be in the same cluster. Since the number of possible pairs in a database is $n(n-1)/2$, it grows as O($n^2$) with the number of records $n$. Since $n$ can be millions of authors in some databases such as PubMed, the AND algorithm needs a blocking method to scale \cite{christen2012survey}. A list of candidate pairs is generated by blocking, and only the pairs on the list are considered for clustering. A good blocking method should have a balance between efficiency and completeness. Efficiency means to minimize the number of pairs to be considered, while completeness implies ensuring coreferent pairs remains after blocking.
	
\begin{figure}[!t]
		\centering
		\includegraphics[clip=true, width=0.7\hsize]{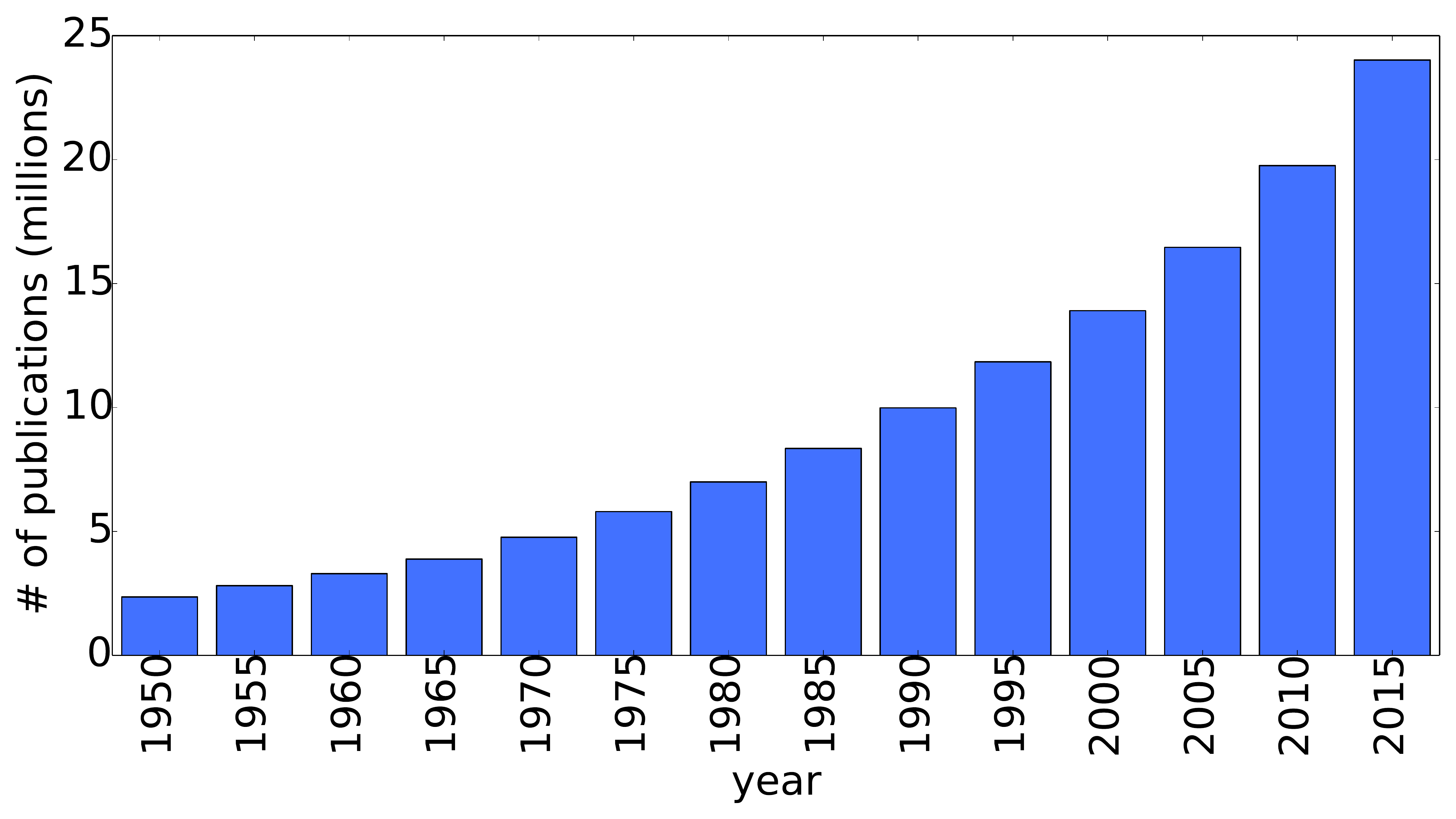}
		\caption{Cumulative Number of Publication Records in the PubMed Database}
		\label{fig:medline_stat}
	\end{figure}

	Blocking is made up of blocking predicates. Each predicate is a logical binary function that selects a set of records based on a combination of an attribute (blocking key) and a similarity criterion. One example can be “exact match of the last name”. A simple but effective way of blocking is selecting manually, with respect to the data characteristics. Most recent work on large-scale AND uses a heuristic with \emph{“initial match of first name”} and \emph{“exact match of last name”} \cite{torvik2009author,liu2014author,levin2012citation,khabsa2014large,kim2016random,kim2016inventor}. 
	Although there is reasonable completeness, it can be problematic when the database is extremely large, such as CiteSeerX (10.1M publications, 32.0M authors), PubMed (24.4M publications, 88.0M authors), and Web of Science (45.3M publications, 162.6M authors)\footnote{Those numbers are measured at the end of 2016.}.
	Table \ref{tab:medlineblock} shows the blocking result on PubMed using the heuristic. Result shows most of the block sizes are less than 100 names, but a few blocks are extremely large. Since the number of pairs grows quadratically, those few blocks can dominate the calculation time. This imbalance of the block size is due to the popularity of certain surnames, especially for Asian names. 
	To make matters worse, this problem increases as time goes by, since the growth rates of the publication records are rapidly increasing. Figure \ref{fig:medline_stat} shows PubMed's cumulative number of publication records.
	
	To improve the blocking, some \cite{bilenko2006adaptive,michelson2006learning,cao2011leveraging,kejriwal2013unsupervised,das2012automatic,fisher2015clustering} have proposed learning it. These approaches can be categorized into two different types. One is disjoint blocking, where each block is separated so that no record belongs to multiple blocks. 
	\cite{das2012automatic,fisher2015clustering} belong to this category. The other is non-disjoint blocking, where some blocks have shared records. 
	\cite{bilenko2006adaptive,michelson2006learning,cao2011leveraging,kejriwal2013unsupervised} are some examples. 
	Each has advantages, disjoint blocking makes the clustering step easily parallelized; thus, each block produced can be independently clustered. Non-disjoint blocking produces smaller blocks, since it uses both disjunction and conjunction, and also has more degrees of freedom to select the similarity criterion. 
	
	Here, we first propose to learn a non-disjoint blocking with a conjunctive normal form (CNF). Also, we propose to extend the method to produce disjoint blocks. Our main contribution can be summarized as below:
	
	\begin{itemize}
		\item We propose a CNF blocking inspired from CNF learning \cite{mooney1995encouraging}. We show that CNF blocking reduces more pairs compared to DNF blocking to achieve high pairs completeness, in the domain of scholarly data. Furthermore, since early rejection of pairs is available, processing time is faster.
		\item To take advantage of disjoint blocking, we extend the method to produce disjoint blocks, so that the clustering step of the AND process can be easily parallelized.
		\item Gain function is used to find the best term to add in each step for learning blocking function. We compare different gain functions introduced in previous work.
	\end{itemize}
	
	
	The paper is organized as follows. In the next session, we discuss previous work. This is followed with the problem definition. Next, we describe learning of CNF blocking and how to make use of CNF blocking while ensuring production of disjoint blocks. After that, we evaluate our methods with a PubMed evaluation dataset. Finally, the last section is a summary of our work including proposals for future directions.

	
	\begin{table}[!t]
		\caption{Block Size Distribution of PubMed with the Simple Heuristic}
		\label{tab:medlineblock}
		\centering
		\resizebox{0.5\hsize}{!}{\begin{tabular}{l||r|r} \hline 
				Block Size&Frequency&Percentage\\ \hline
				$2 \leq n < 10$& 1,586,677 & 59.91\%\\
				$10 \leq n < 100$& 910,272 & 34.37\% \\
				$100 \leq n < 1000$& 144,361 & 5.45\% \\
				$1000 \leq n < 10000$ & 6,998 & 0.26\% \\
				$10000 \leq n < 50000$ &184 & 0.01\% \\ 
				$n \geq 50000 $ & 9 & $<$ 0.01\%\\ \hline
				Total & 2,648,501 & 100.0 \%\\ \hline
		\end{tabular}}
	\end{table}
	
	
	\section{Related Work} \label{section:related}
	Blocking has been widely studied for record linkage and disambiguation. Standard blocking is the simplest but most widely used method\cite{fellegi1969theory}. It is done by considering only pairs that meet all blocking predicates. Each blocking predicate is consistent with a selection of a blocking key and similarity criterion. Another is the sorted neighborhood approach \cite{hernandez1995merge}. It sorts the data by a certain blocking predicate, then for each record, it forms pairs with those records within a window for further processing. Yan et al. \cite{yan2007adaptive} further improved this method to select the size of the window adaptively. Aizawa and Oyama \cite{aizawa2005fast} introduced a suffix array-based indexing, which uses an inverted index of suffixes to generate candidate pairs. Each record generates pairs with records with at least one shared suffix. There is also canopy clustering \cite{mccallum2000efficient}, which generates blocks by clustering with a simple similarity measure and two thresholds, namely loose similarity and tight similarity. While generating a cluster, all records within the loose similarity will be inserted into the cluster, but only those within tight similarity will be removed from the set of candidate records, so that the algorithm generates overlapping clusters.
	
	Most of those methods were compared in two recent surveys by Christen \cite{christen2012survey} and Papadakis et al. \cite{papadakis2016comparative}. Those surveys show that no clear winners are among methods, and proper parameter tuning is required for a specific task. Thus, in this paper, we mainly focus on optimizing the blocking function for standard blocking, because of its simplicity to use and small computational overhead of applying it. Another benefit is blocks produced by a standard blocking are easy to parallelize for subsequent processing, such as clustering, as long as those blocks are mutually exclusive.
	
	Much work optimized the blocking function for standard blocking. The blocking function is typically presented with a logical formula with blocking predicates. Two of which on learning a disjunctive normal form (DNF) blocking \cite{bilenko2006adaptive,michelson2006learning} were published in the same year. Making use of manually labeled record pairs, they used a sequential covering algorithm to find the optimal blocking predicates in a greedy manner. 
	Additional unlabeled data was used to estimate the reduction ratio of their cost function \cite{cao2011leveraging} while an unsupervised algorithm was used to automatically generate labeled pairs with rule-based heuristics used to learn DNF blocking \cite{kejriwal2013unsupervised}.
	
	All the work above proposed to learn DNF blocking, which is a disjunction (\emph{OR}) of conjunctions (\emph{AND}). They produce non-disjoint blocks because of the logical \emph{OR} terms. On the other hand, other work learns the blocking function with a pure conjunction, to ensure to generate disjoint blocks. Das et al. \cite{das2012automatic} learns a conjunctive blocking tree, which has different blocking predicates for each branch on the tree. Fisher et al. \cite{fisher2015clustering} produces blocks with respect to a size restriction, by generating candidate blocks with a list of predefined blocking predicates and then performs a merge and split step to generate the block with the desired size.
	
	Our work proposes a method for learning a non-disjoint blocking function in a conjunctive normal form (CNF), and later extend it to ensure generating disjoint blocks, which is not previously studied in this field. Our method is inspired from the CNF learner proposed in \cite{mooney1995encouraging}, which is the idea that CNF can be a logical dual of DNF. 
	
	\section{Problem Definition} \label{section:problem_definition}
	
	Our work tackles the same problem with the DNF blocking \cite{bilenko2006adaptive,michelson2006learning}. Let $R=\{r_1, r_2, \cdots, r_n\}$ be the set of records in the database, where $n$ is the number of records. Each record $r$ has $k$ attributes, and $A$ be the attribute set $A=\{a_1. a_2, \cdots, a_k\}$. A blocking predicate $p$ is a combination of an attribute $a$ and a similarity function $s$ defined to $a$. An example of $s$ is exact string match of $a$. It is a logical binary function applied to each pair of records, so $p(r_x, r_y) = \{0,1\}$, where $r_x, r_y \in R$. Blocking function $f$ is a boolean logic formula consisting with blocking predicates $p_1, p_2, \cdots, p_n$, and each predicate is connected with either conjunction $\wedge$ or disjunction $\vee$. An example is $f_{example} = (p_1 \wedge p_2) \vee p_3$. Since it is made up of blocking predicates, $f(r_x,r_y) = \{0,1\}$ for all $r_x, r_y \in R$. 
	
	The goal is to find an optimal blocking function $f^*$ that covers a minimum number record pairs while missing up to $\varepsilon$ of total number of matching record pairs. To formalize it,
	\begin{align}\label{eq:formulize}
	\begin{split}
	f^* = \argmin_f\sum_{(r_x, r_y)\in R} f(r_x, r_y) \\
	\text{s.t. } \sum_{(r_i, r_j)\in R^+} f(r_i, r_j) \geq (1-\varepsilon) \times \lvert R^+ \rvert
	\end{split}
	\end{align}
	where $R^+$ is set of matching record pairs.
	
	\section{Learning Blocking Function} \label{section:method}
	In this section, we first review the DNF blocking \cite{bilenko2006adaptive,michelson2006learning}. Then we introduce our CNF blocking, which can be implemented with a small modification from the DNF blocking. Then, we describe several gain functions. It is used to select an optimal predicate term in each step for learning CNF and DNF. Finally, we discuss an extension to ensure the production of disjunctive blocks for easy parallelization.
	
	\subsection{DNF blocking}
	\begin{algorithm}[!h]
		\caption{DNF Blocking}
		\label{algorithm:DNF}
		\begin{algorithmic}[1]
			\Function{LearnConjTerms}{$L,P,p,k$}
			\State Let $Pos$ be set of positive samples in $L$
			\State Let $Neg$ be set of negative samples in $L$
			\State $Terms \gets \{p\}$
			\State $CurTerm \gets p$
			\State $i \gets 1$
			\While{$i < k$}
			\State Find $p_i \in P$ that maximizes gain function \Call{CalcGain}{$Pos, Neg, CurTerm \wedge p_i$}
			\State $CurTerm \gets CurTerm \wedge p_i$
			\State Add $CurTerm$ to $Terms$
			\State $i \gets i+1$
			\EndWhile
			\State \Return $Terms$
			\EndFunction
			
			\State
			
			\Function{LearnDNF}{$L, P, k$} 
			\State $CandTerms \gets \phi$
			\For{$p \in P$}
			\State $Terms\gets$\Call{LearnConjTerms}{$L, P, p, k$}
			\State $CandTerms \gets CandTerms \cup Terms$
			\EndFor
			\State Let $Pos$ be set of positive samples in $L$
			\State Let $Neg$ be set of negative samples in $L$
			\State $DNF \gets \phi$
			\While{$\lvert Pos \rvert > \varepsilon \times \lvert Pos \rvert$}
			\State Find $T \in CandTerms$ that maximizes gain function \Call{CalcGain}{$Pos, Neg, T$} 
			\If{\Call{CalcGain}{$Pos, Neg, t$} $ > 0$}
			\State $DNF \gets DNF \vee T$ 
			\State Let $PosCov$ be all $l \in Pos$ that satisfies $T$
			\State Let $NegCov$ be all $l \in Neg$ that satisfies $T$
			\State $Pos \gets Pos - PosCov$ 
			\State $Neg \gets Neg - NegCov$ 
			\Else
			\State break loop
			\EndIf
			\EndWhile
			\State \Return $DNF$
			\EndFunction
			
			%
			%
		\end{algorithmic}
	\end{algorithm}
	
	DNF blocking is originally proposed by two parallel work \cite{bilenko2006adaptive,michelson2006learning} in the same year. Although some details are different, the main idea is similar. Given labeled pairs, they attempt to learn the blocking function in the form of DNF, which is the disjunction (logical \emph{OR}) of conjunction (logical \emph{$AND$}) terms. 
	Learning DNFs are known to be a NP-hard problem \cite{bilenko2006adaptive}, they proposed an approximation algorithm to learn $k$-DNF blocking using a sequential covering algorithm. $k$-DNF means each conjunction term has, at most, $k$ predicates.  
	
	Algorithm \ref{algorithm:DNF} shows the process of DNF blocking. Function \textsc{LearnDNF} in line 16-39 is the main function of the algorithm. It gets 3 inputs, $L$ is labeled sample pairs, $P$ is blocking predicates, $k$ is the parameter of maximum predicates considered in each conjunction term.
	
	First, to reduce the computation, the algorithm selects a set of candidate conjunction terms with at most $k$ predicates. It is not practical to use all possible combinations, since the time complexity will be exponential to $k$. For each predicate $p$, it generates $k$ candidate conjunction terms with the function \textsc{LearnConjTerms} in line 1-14. It iteratively selects a predicate $p_i$, which has the best gain value when it is added into the conjunction term selected from previous step (line 8). Gain value is calculated by the function \textsc{CalcGain}, there are several different metrics to calculate it. We discuss them in detail in later section.
	
	Using the candidate terms, the algorithm learns a DNF blocking function by running a sequential covering algorithm (line 26-35). In each iteration, it sequentially selects a conjunction term $Term$ from the set of candidate conjunction terms $Terms$ that has the maximum gain value, and attach it with logical \emph{OR} to the DNF term. Gain is again calculated with the function \textsc{CalcGain}. Then all positive and negative samples covered by $Term$ are removed. This process repeats until it covers the desired minimum amount of positive samples, or there is no additional candidate term that produces a positive gain value. 
	
	\subsection{CNF blocking}
	\begin{algorithm}[!h]
		\caption{CNF Blocking}
		\label{algorithm:CNF}
		\begin{algorithmic}[1]
			\Function{LearnNegConjTerms}{$L,NegP,p,k$}
			\State Let $Pos$ be set of positive samples in $L$
			\State Let $Neg$ be set of negative samples in $L$
			\State $Terms \gets \{p\}$
			\State $CurTerm \gets p$
			\State $i \gets 1$
			\While{$i < k$}
			\State Find $p_i \in NegP$ that maximizes gain function \Call{CalcNegGain}{$Pos, Neg, CurTerm \wedge p_i$}
			\State $CurTerm \gets CurTerm \wedge p_i$
			\State Add $CurTerm$ to $Terms$
			\State $i \gets i+1$
			\EndWhile
			\State \Return $Terms$
			\EndFunction
			
			\State
			
			\Function{LearnCNF}{$L, P, k$}
			\State $CandTerms \gets \phi$
			\State Let $NegP$ is negation of each $p \in P$
			\For{$p \in NegP$}
			\State $Terms\gets$\Call{LearnNegConjTerms}{$L, NegP, p, k$}
			\State $CandTerms \gets CandTerms \cup Terms$
			\EndFor
			\State Let $Pos$ be set of positive samples in $L$
			\State Let $Neg$ be set of negative samples in $L$
			\State $NegDNF \gets \phi $
			\While{$\lvert Pos \rvert >  (1-\varepsilon) \times \lvert Pos \rvert$}
			\State Find $T \in CandNegTerms$ that maximizes gain function \Call{CalcNegGain}{$Pos, Neg, Term$} 
			\If{\Call{CalcNegGain}{$Pos, Neg, T$} $ > 0$}
			\State $NegDNF \gets NegDNF \vee T$ 
			\State Let $PosCov$ be all $l$ in $Pos$ that satisfies $T$
			\State Let $NegCov$ be all $l$ in $Neg$ that satisfies $T$
			\State $Pos \gets Pos - PosCov$ 
			\State $Neg \gets Neg - NegCov$ 
			\Else
			\State break loop
			\EndIf
			\EndWhile
			\State $CNF \gets \neg (NegDNF)$ 
			\State \Return $CNF$
			\EndFunction
			
			%
		\end{algorithmic}
	\end{algorithm}
	CNF blocking can be learned with a small modification from the DNF blocking algorithm. First, we review some basic boolean algebra to understand the relation between CNF and DNF. CNF can be presented as the entire negation of a corresponding DNF and vice versa, using De Morgan's laws. De Morgan's laws are as follows:
	\begin{align}
	\neg (A \wedge B) \leftrightarrow (\neg A) \vee (\neg B) \label{eq:de_morgan1} \\
	\neg (A \vee B) \leftrightarrow (\neg A) \wedge (\neg B) \label{eq:de_morgan2} 
	\end{align}
	
	For example, let's assume that we have a DNF formula $\neg A \vee (\neg B \wedge \neg C)$. The negation of the formula is a CNF formula by using (\ref{eq:de_morgan1}) and (\ref{eq:de_morgan2}):
	\begin{align*}\label{eq:DNF_to_CNF} 
	\neg(\neg A \vee (\neg B \wedge \neg C))  
	&= A \wedge \neg (\neg B \wedge \neg C) \\
	&= A \wedge (B \vee C)
	\end{align*} 
	
	Using this fact, Mooney proposed an approximate CNF learning \cite{mooney1995encouraging}, which is a logical dual of DNF learning. Inspired from it, we present our CNF blocking method, which is a logical dual of the DNF blocking. 
	
	Algorithm \ref{algorithm:CNF} shows the proposed CNF blocking. The algorithm has similar structure as Algorithm \ref{algorithm:DNF}. Instead of running a sequential covering algorithm to cover all positive samples, CNF blocking first tries to cover all negative samples using negated blocking predicates. In other words, we learn a DNF formula that is consistent with a negated predicate, which we call negated DNF ($NegDNF$ in Algorithm \ref{algorithm:CNF}). $NegP$ is the negation of each predicate $p$ in $P$. Main function \textsc{LearnCNF} gets 3 inputs as \textsc{LearnDNF} in Algorithm \ref{algorithm:DNF}. $L$ is labeled sample pairs, $P$ is blocking predicates, $k$ is the parameter of maximum predicates in each term.
	
	The algorithm runs as follows. First, it generates a set of negated candidate conjunction term $Terms$ from all $p$ in $NegP$ (line 16). We use a dual of the original gain function, \textsc{CalcNegGain} to select a predicate for generating a negated candidate conjunction. Then, as in the DNF blocking, it runs the sequential covering algorithm to learn the negated DNF formula (line 27-38), which iteratively adds a negated conjunction term until it covers the desired amount of samples. We select a negated conjunction term with a dual of the original gain function, \textsc{CalcNegGain}. Also, note that the termination condition of the loop (line 27) is when $\varepsilon$ of total positive samples are covered with the learned $NegDNF$, to ensure we miss less than $\varepsilon$ of the total number of positive samples in the final CNF formula. After we get the final $NegDNF$, we negate it to get the desired CNF. 
	
	\subsection{Gain Function} \label{subsection:gain}
	Gain function estimates the benefit of adding a specific term to the current learned formula for DNF / CNF blocking. It is used in two different places in the algorithm; one is when we choose the conjunction candidates (line 8 in both Algorithm \ref{algorithm:DNF} and \ref{algorithm:CNF}), the other is when we choose a term from the candidates in each iteration (line 27 in Algorithm \ref{algorithm:DNF}, line 28 in Algorithm \ref{algorithm:CNF}). Previous methods \cite{mooney1995encouraging,bilenko2006adaptive,michelson2006learning} proposed a different gain function, here we describe the original function for DNF blocking and its dual to use for our CNF blocking, and we compare the results in the experiments section. $P$, $N$ is the total number of positive / negative samples. In addition, $p$, $n$ is the number of remaining positive / negative samples covered by the term. 
	
	\subsubsection{Information Gain}
	It is originally from Mooney's DNF and CNF learner \cite{mooney1995encouraging}. Original gain function for DNF learner can be calculated as 
	
	\begin{equation}
	gain_{DNF} =  p \times \bigg[\log\bigg(\frac{p}{p+n}\bigg) - \log\bigg(\frac{P}{P+N}\bigg)\bigg]
	\end{equation} 
	The gain function of CNF learner can be calculated in the same way, in this case since we cover negative samples, 
	\begin{equation}
	gain_{CNF} =  n \times \bigg[\log\bigg(\frac{n}{n+p}\bigg) - \log\bigg(\frac{N}{N+P}\bigg)\bigg]
	\end{equation} 
	
	\subsubsection{Ratio Between Positive and Negative Samples Covered} \label{subsection:PosNegRatio}
	Bilenko et al. \cite{bilenko2006adaptive} used this for their DNF blocking. It calculates the ratio between the number of positives and the number of negatives covered.
	\begingroup
	\begin{align} 
	gain_{DNF} = \frac{p}{n} \label{eq:ratio1}\\
	gain_{CNF} = \frac{n}{p} \label{eq:ratio2}
	\end{align}
	\endgroup
	
	\subsubsection{Reduction Ratio}
	Michaelson and Knoblock \cite{michelson2006learning} picked terms with the maximum reduction ratio (RR). In addition, they filter out all terms that has with pairwise completeness (PC) below threshold $t$. This can be presented as:
	
	\begin{equation}
	gain_{DNF} = 
	\begin{cases}
	\frac{p+n}{P+N} & \text{if $\frac{p}{P} > t$}\\
	0 & \text{otherwise}
	\end{cases}
	\end{equation}
	\begin{equation}
	gain_{CNF} = 
	\begin{cases}
	\frac{p+n}{P+N} & \text{if $\frac{n}{N} > t$}\\
	0 & \text{otherwise}
	\end{cases}
	\end{equation}
	
	\subsection{Learning Disjoint Blocks} \label{subsection:disjoint}
	Blocking methods can be categorized into non-disjoint and disjoint. A blocking method is disjoint if all records are separated into mutual exclusive blocks by applying the method. 
	A blocking function is disjoint if and only if it suffices the following conditions: 1) it only consists of pure conjunction (logical \emph{AND}), 2) all predicates use non-relative similarity measures. That is, measures comparing absolute value of blocking key, e.g. exact match of first $n$ characters. 
	
	DNF and CNF blocking are both forms of non-disjoint blocking due to the first condition. A disjoint blocking has the advantage that parallelization can be performed easily after applying the blocking method, by running processes for each blocks separately. A weakness of this form is that an ordinary disjoint method tends to have larger blocks, because its formula cannot use logical \emph{OR}, and only non-relative similarity measures can be used for candidate predicates. 
	
	We introduce a simple extension to ensure our CNF blocking to produce disjoint blocks (Algorithm \ref{algorithm:disjoint CNF}). This is carried out by first producing two blocking functions. The first function learns a blocking function with only conjunctions. This can be performed by running our CNF blocking method with $k=1$ and using a set of predicates with non-relative similarity measures $P_{disjoint}$, since $1$-CNF equals to pure conjunction. Then, we learn a CNF blocking with our ordinary $k$-CNF method with the whole set of predicates $P_{full}$, for pairs remaining after applying $1$-CNF . 
	
	After learning them, we first apply the $1$-CNF to the whole database to produce disjoint blocks. Then for each block, we apply the second $k$-CNF blocking function to identify pairs for further processing (clustering).  Thus, we filter out pairs that are not consistent with $k$-CNF, considering them as non-matched pairs. This is similar to the filtering method from Gu and Baxter \cite{gu2004adaptive} and Khabsa et al. \cite{khabsa2015online}. The difference is that we learn those instead of choosing heuristically. 
	
	This method still produces CNF since it combines conjunction terms and $k$-CNF with logical \emph{AND}. The first part consisting of pure conjunction ensures the production of non-disjoint blocks.
	
	\begin{algorithm}[!t]
		\caption{Disjoint CNF Blocking}
		\label{algorithm:disjoint CNF}
		\begin{algorithmic}[1]
			\Function{DisjointCNF}{$L, P_{disjoint}, P_{full}, k$}
			\State $Conj \gets \Call{LearnCNF}{L, P_{disjoint}, 1}$
			\State Let $L^{\prime}$ be set of $l \in L$ satisfies $Conj$
			\State $CNF \gets \Call{LearnCNF}{L_{remain},  P_{full}, k}$
			\State $Blocks \gets$ Apply $Conj$ to whole data
			\For{$Block \in Blocks$}
			\State Let $L^{\prime\prime}$ be $l \in Block$ that satisfies $CNF$
			\State Consider pairs in $L^{\prime\prime}$ only for clustering
			\EndFor
			\EndFunction
		\end{algorithmic}
	\end{algorithm}
	
	\section{Experiments} 
	\label{section:result}
	\subsection{Benchmark Dataset}
	\begin{table}
		\caption{Summary of PubMed Benchmark Dataset}
		\label{tab:benchmark_stat}
		\centering	
		\resizebox{0.7\hsize}{!}{\begin{tabular}{r|r|r|r} \hline 
				\# Authors & \# Mentions & \# Total Pairs & \# Matched Pairs \\ \hline
				214 & 3,964 & 7,854,666 & 51,052 \\ \hline
		\end{tabular}}
	\end{table}
	
	We use the PubMed for the evaluation. PubMed is a public large-scale scholarly database maintained by the National Center for Biotechnology Information (NCBI) at the National Library of Medicine (NLM). 
	We use NIH principal investigator (PI) data for evaluation, which include PI IDs and corresponding publications. We randomly picked 10 names from the most frequent ones in the dataset and verified that all publications belong to each PI. The set of names include C* Lee, J* Chen, J* Smith, M* Johnson, M* Miller, R* Jones, S* Kim, X* Yang, Y* Li, Y* Wang, where C* means any name starts with C. Table \ref{tab:benchmark_stat} shows the statistics of the dataset. Experiments are done with 2-fold cross validation.
	
	
	
	
	\subsection{Methodology}
	\subsubsection{Evaluation Metrics}
	We evaluate our CNF blocking with reduction ratio (RR), pairs completeness (PC), and F-measure. These metrics are often used to evaluate blocking methods. Those metrics can be calculated as follows:
	\begingroup
	\begin{align} 
	RR &= 1 - \frac{p+n}{P+N} \\
	PC &= \frac{p}{P} \\
	F &= \frac{2 \times RR \times PC}{RR + PC}
	\end{align}
	\endgroup
	where $P$, $N$ are the numbers of positive/negative samples, and $p$, $n$ are the numbers of positive/negative samples covered with the blocking function. RR measures the efficiency of the blocking function, PC measures the quality of the blocking function. F is the harmonic mean of RR and PC.
	
	\subsubsection{Blocking Predicates Used}
	\label{subsection:comp}
	We define two different sets of blocking predicates. As we discussed in the previous section, disjoint blocking requires the use of predicates with non-relative similarity measures (e.g., exact match) to ensure blocks are mutually exclusive. On the other hand, non-disjoint blocking has more degrees of freedom to include relative similarity measures (e.g., TF-IDF cosine distance). The blocking predicates set for disjoint and non-disjoint blocking are described in Table \ref{tab:predicates1} and Table \ref{tab:predicates2}, respectively.
	
	We observed an important characteristic of the data: some attributes are empty. For example, 92.2$\%$ include year, 19.9$\%$ have affiliation, 54.5$\%$ of the author mentions have only initials for the first name. To deal with it better, we add $compatible$ for those blocking keys.
	
	Below is a brief explanation for each similarity criterion.
	
	\begin{itemize}
		\item{$exact$}: Exact match.
		\item{$first(n), last(n)$}: First/Last $n$ character match, where $n$ is an integer. We check $\{1, 3, 5, 7\}$ for name attributes.
		\item{$order$}: Assigns $True$ if both records are first authors, last authors, or non-first and non-last authors.
		\item{$digit(n)$}: First $n$ digit match. We check $\{1, 2, 3\}$ for year.
		\item{$compatible$}: $True$ if at least one of the records are empty (Eq. \ref{eq:comp}). If the key is name, it also checks if the initial matches if one of the records has only initial (Eq. \ref{eq:name_comp}).
		\begin{equation}
		\label{eq:comp}
		\resizebox{0.65\hsize}{!}{
			$compatible(A,B) = 
			\begin{cases}
			True & \text{if at least one is empty} \\
			exact(A,B) & \text{otherwise}\\
			\end{cases}$}
		\end{equation}
		\begin{equation}
		\label{eq:name_comp}
		\resizebox{0.65\hsize}{!}{
			$compatible(A,B) = 
			\begin{cases}
			True & \text{if at least one is empty} \\
			exact(A,B) & \text{if both are full name}\\
			first1(A,B) & \text{otherwise}\\
			\end{cases}$}
		\end{equation}
		\item{$cos$}: Cosine distance of TF-IDF bag-of-words vector. We check with threshold $\{0.2, 0.4, 0.6, 0.8\}$.
		\item {$diff$}: Year difference. We use the threshold $\{2, 5, 10\}$.
	\end{itemize} 
	
	\begin{table}[!t]
		\caption{Blocking Predicates Used for Learning Disjoint Blocking Function}
		\label{tab:predicates1}
		\centering	
		\resizebox{0.43\hsize}{!}{\begin{tabular}{l|l} \hline 
				Blocking Key & Similarity Criterion\\ \hline
				First Name & $exact, first(n), last(n)$ \\
				Last Name & $exact, first(n), last(n)$  \\
				Middle Name & $exact, first(n), last(n)$ \\
				Affiliation & $exact$ \\
				Order & $order$ \\ 
				Year & $exact, digit(n)$ \\ 
				Venue & $exact$ \\ \hline
		\end{tabular}}
	\end{table}
	\begin{table}[!t]
		\caption{Blocking Predicates Used for Learning Non-Disjoint Blocking Function}
		\label{tab:predicates2}
		\centering	
		\resizebox{0.55\hsize}{!}{\begin{tabular}{l|l} \hline 
				Blocking Key & Similarity Criterion\\ \hline
				First Name & $exact, first(n), last(n), compatible$ \\
				Last Name & $exact, first(n), last(n), compatible$  \\
				Middle Name & $exact, first(n), last(n), compatible$ \\
				Title & $cos$ \\
				Affiliation & $exact, cos, compatible$ \\
				Coauthor & $cos$ \\
				Order & $order$ \\ 
				Year & $exact, digit, diff$ \\ 
				Venue & $exact, cos$ \\ \hline
		\end{tabular}}
	\end{table}
	
	\subsubsection{Parameter Setting}
	
	$\varepsilon$ is used to vary the PC; we tested values inside $[0,1]$ to get the PC--RR curve. $k$ is selected experimentally to calculate the maximum reachable F-measure. The result was 0.9458, 0.9531, 0.9540, 0.9535 for $k=1, 2, 3, 4$ respectively. For $k>5$, the result was the same as with $k=4$, which means no terms were selected in more than four predicates. From the result, $k=3$ had the highest F so we use it for further experiments.
	
	\subsection{Experimentation Result}

	\subsubsection{Gain Function}
	\begin{figure}[!t]
		\centering
		\includegraphics[clip=true, width=0.65\hsize]{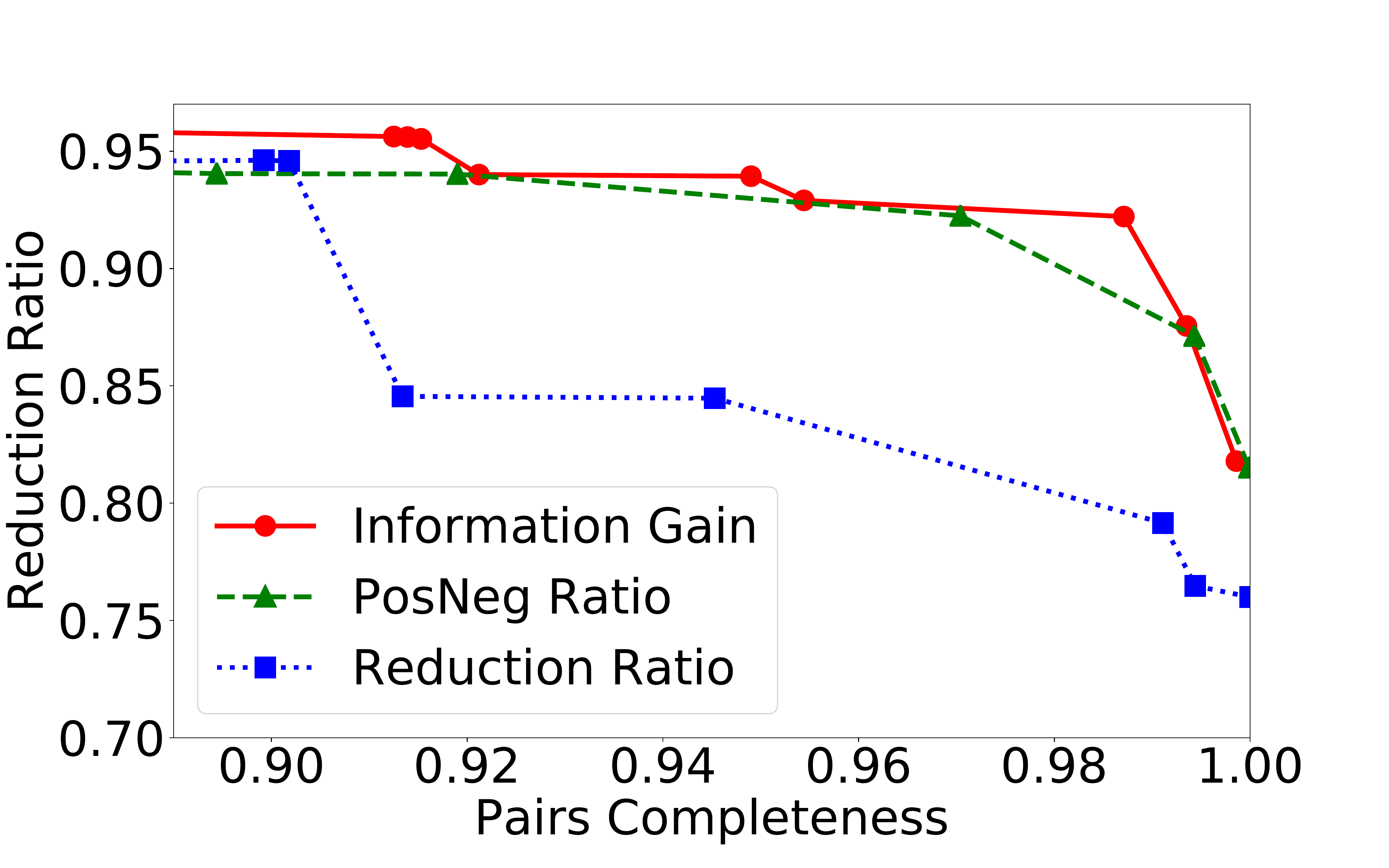}
		\caption{Comparison of Gain Functions. PosNeg Ratio is the ratio between positive and negative samples covered.}
		\label{fig:gain}
	\end{figure}
	We tested three different gain functions introduced in the previous section. Figure \ref{fig:gain} shows the PC--RR curve generated by testing various $\varepsilon$ values. Blocking usually requires high PC, so that we do not lose matched pairs after it is applied. So, we focused on experimenting for high PC values. As we can see from the results, information gain has highest RR overall. Thus, we use it as the gain function for the rest of the experiments.
	
	\subsubsection{Non-disjoint CNF Blocking} \label{subsection:main_comp}
	\begin{figure}[!t]
		\centering
		\includegraphics[clip=true, width=0.65\hsize]{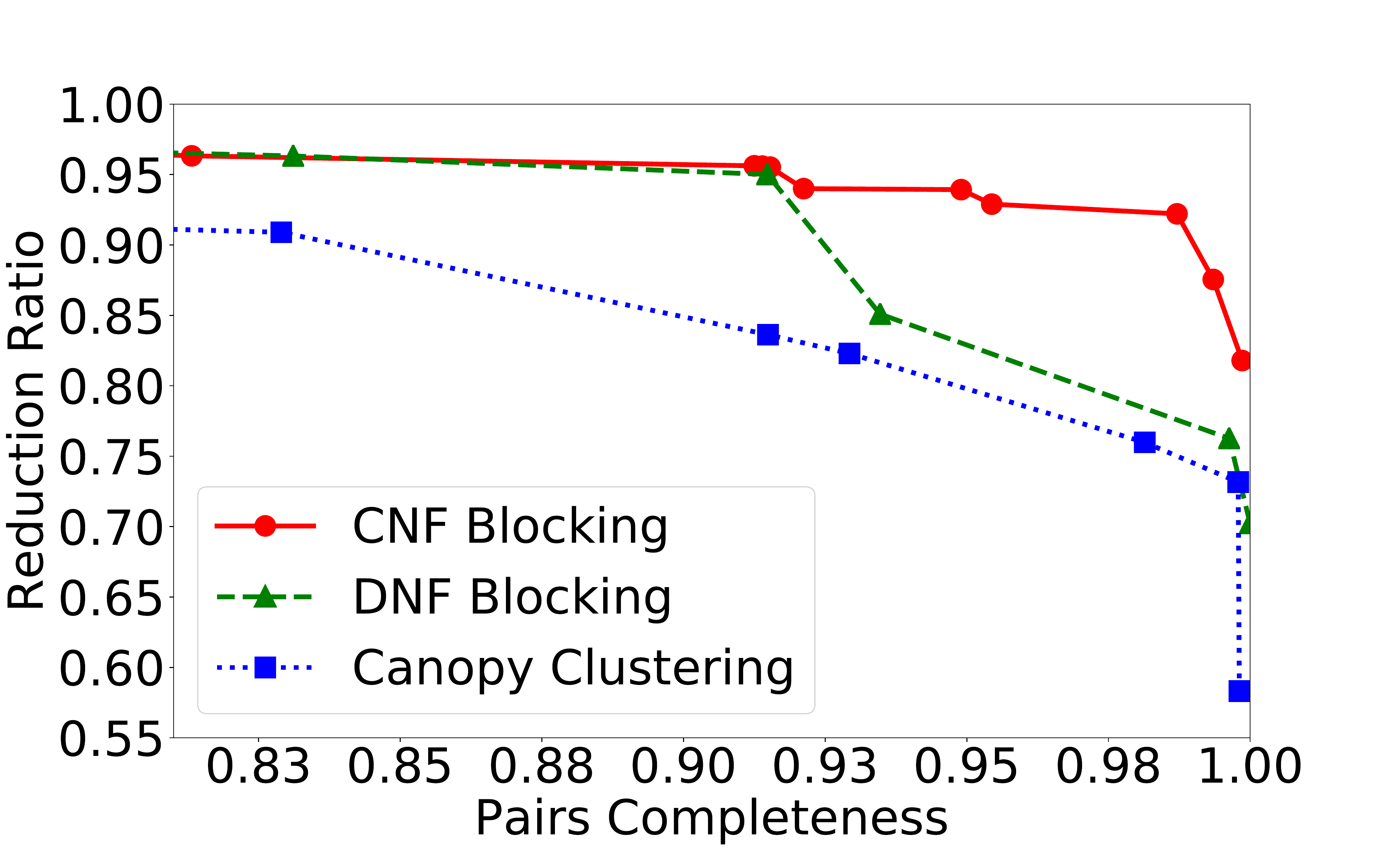}
		\caption{Comparison of Non-disjoint Blocking Methods}
		\label{fig:medline}
	\end{figure}
	
	We compare the non-disjoint CNF blocking method with the DNF blocking and canopy clustering \cite{mccallum2000efficient}. We use a supervised method to train DNF \cite{bilenko2006adaptive,michelson2006learning} to take advantage of training data. We used the set of Jaro--Winkler distance of attributes for canopy clustering.
	
	Figure \ref{fig:medline} shows the PC--RR curve for each method. Both CNF and DNF were better than canopy clustering, as was shown in Bilenko et al. \cite{bilenko2006adaptive}. CNF and DNF results are comparable except for high PC ($\verb+>+ 0.9$) values. From the result, concepts in the domain of scholarly data can be better represented with CNFs. For the AND problem, the blocking requires high PC so that positive pairs are not excluded in the clustering process, so CNF is preferred.
	
	
	\begin{table}[!t]
		\caption{Processing Time of Non-disjoint Blocking Methods}
		\label{tab:process_nondisjoint}
		\centering	
		\resizebox{0.7\hsize}{!}{\begin{tabular}{l|c|c} \hline 
				Method & Training Time (s) & Blocking Time (s) \\ \hline
				CNF Blocking & 147.36 & 1.39 \\ 
				DNF Blocking & 564.25 & 2.09 \\
				Canopy Clustering & N/A & 0.44 \\ \hline
		\end{tabular}}
	\end{table}
	
	Another advantage of using CNF is the processing time. Table \ref{tab:process_nondisjoint} compares the training time and blocking time at PC=0.99, excluding calculation of string distances. The computation time of CNF is significantly reduced compared to DNF. This is because CNF is composed with conjunction of disjunction terms, it can quickly reject pairs that are not consistent with any terms. On the other hand, DNF consists of disjunction of conjunction terms, so each pair should check all terms to make the final decision. Canopy clustering is the fastest, but it degrades RR significantly in high PC setting. Learned CNF is also simpler than DNF. Learned CNF at this level is as below (fn, mn, ln is first, middle, last name respectively):  \\
	\emph{\{(fn,first(5))$\vee$(fn,compatible)$\vee$(coauth,cos(0.8))\} $\wedge$\\
		\{(ln,exact)\} $\wedge$ \{(mn,compatible)\} $\wedge$\\
		\{((fn,first(3))$\vee$(fn,compatible))\}} \\
	
	And learned DNF is: \\
	\emph{\{((coauth,cos(0.8))$\wedge$(ln,exact)$\wedge$(mn,compatible))\} $\vee$\\
		\{((venue,cos(0.4))$\wedge$(mn,first(1)$\wedge$(fn,compatible)\} $\vee$\\
		\{(fn,compatible)$\wedge$(mn,first(1)$\wedge$(ln,exact)\} $\vee$\\
		\{((venue,cos(0.8))$\wedge$(fn,exact)\} } \\
	
	In addition, we observed that proposed $compatible$ predicate was frequently used in our result. This shows the effectiveness of $compatible$ in dealing with the empty value.
	
	\subsubsection{Disjoint CNF Blocking}
	\begin{figure}[!t]
		\centering
		\includegraphics[clip=true, width=0.65\hsize]{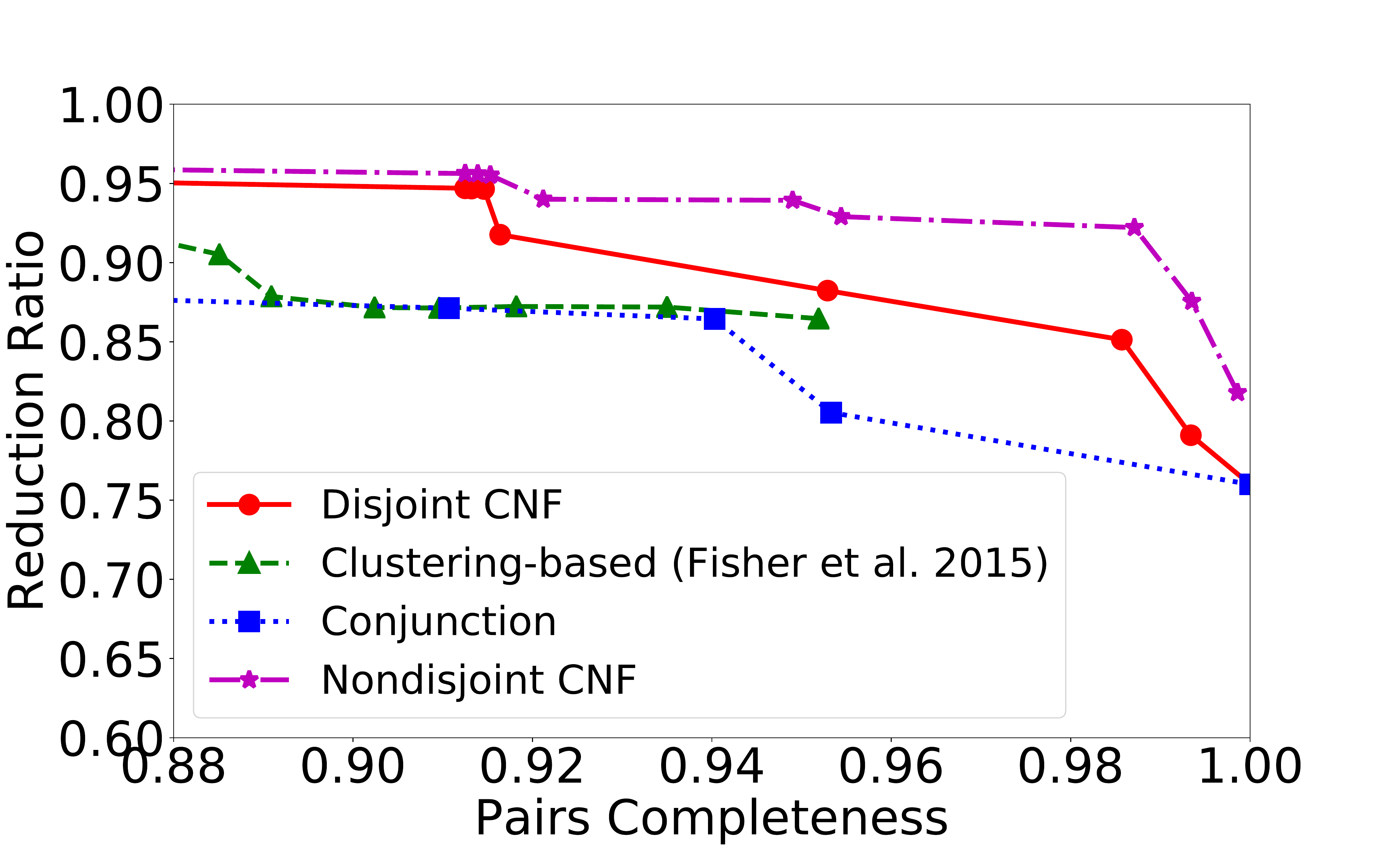}
		\caption{Comparison of Disjoint Blocking Methods}
		\label{fig:hybrid}
	\end{figure}
	We evaluate our extension to make disjoint blocks with the CNF blocking. We compare the blocking learned with a pure conjunction, our proposed method, and the method of Fisher et al. \cite{fisher2015clustering}. 
	
	Figure \ref{fig:hybrid} shows the RR--PC curve for each method. We also plot the original non-disjoint CNF blocking for comparison. We can see that our proposed disjoint CNF blocking is the winner among non-disjoint methods. Fisher's method produced nearly uniform-sized blocks, but it had a limitation to reach high PC and generally RR was similar to conjunction for the same PC level. Disjoint CNF has some degradation compared with the non-disjoint CNF because it is forced to use pure conjunction and limited predicates in the first part. However, this simple extension can aid to easily parallelize the clustering process. Parallelization is important to make the disambiguation algorithm scale to PubMed-scale scholarly databases \cite{khabsa2014large}.
	
	Processing time for disjoint CNF blocking comparable to the original non-disjoint CNF blocking, it takes 1.57s at PC=0.99. The learned disjoint CNF is: \\
	\emph{\{(fn,first(1))\} $\wedge$ \{(ln,exact)\} $\wedge$\\
		\{(fn,compatible)$\vee$(coauth,cos(0.8))\} $\wedge$ \{(mn,compatible)\}} \\
	
	First two terms are from 1-CNF, and others from 3-CNF learner. We also tested this function to the whole PubMed. 82.17\% of the pairs is reduced using the learned blocking function, and the running time is 10.5 min with 24 threads. 
	
	\section{Conclusion} \label{section:conclusion}
	We show how to learn an efficient blocking function with a conjunctive normal form (CNF) of blocking predicates. Using CNF as a negation of corresponding disjunctive normal form (DNF) of predicates \cite{mooney1995encouraging}, our method is the logical dual of existing DNF blocking methods \cite{bilenko2006adaptive,michelson2006learning}. We find that a learned CNF blocking function reduces more pairs for a large number of target pairs completeness with a faster run time. We devise an extension that ensures that our CNF blocking produces disjoint blocks, so that the clustering process can be easily parallelized. 
	
	Future work could improve our CNF method so it could be used on different levels of blocking functions for each block with a pure conjunction blocking function \cite{das2012automatic}. Instead of using a sequential covering algorithm, the feasibility of using linear programming to find an optimal CNF \cite{su2016learning} could be explored.
	
\section{Acknowledgments}
We gratefully acknowledge partial support from the National Science Foundation.
	
\bibliographystyle{abbrv}
\bibliography{kunho}  

\begin{thebibliography}{10}

\bibitem{aizawa2005fast}
A.~Aizawa and K.~Oyama.
\newblock A fast linkage detection scheme for multi-source information
  integration.
\newblock In {\em International Workshop on Challenges in Web Information
  Retrieval and Integration}, pages 30--39. IEEE, 2005.

\bibitem{bilenko2006adaptive}
M.~Bilenko, B.~Kamath, and R.~J. Mooney.
\newblock Adaptive blocking: Learning to scale up record linkage.
\newblock In {\em Proceedings of the 6th IEEE International Conference on Data
  Mining(ICDM'06)}, pages 87--96, 2006.

\bibitem{cao2011leveraging}
Y.~Cao, Z.~Chen, J.~Zhu, P.~Yue, C.-Y. Lin, and Y.~Yu.
\newblock Leveraging unlabeled data to scale blocking for record linkage.
\newblock In {\em Proceedings of the International Joint Conference on
  Artificial Intelligence (IJCAI)}, volume~22, page 2211, 2011.

\bibitem{christen2012survey}
P.~Christen.
\newblock A survey of indexing techniques for scalable record linkage and
  deduplication.
\newblock {\em IEEE Transactions on Knowledge and Data Engineering (TKDE)},
  24(9):1537--1555, 2012.

\bibitem{das2012automatic}
A.~Das~Sarma, A.~Jain, A.~Machanavajjhala, and P.~Bohannon.
\newblock An automatic blocking mechanism for large-scale de-duplication tasks.
\newblock In {\em Proceedings of the 21st ACM international conference on
  Information and knowledge management (CIKM)}, pages 1055--1064. ACM, 2012.

\bibitem{fellegi1969theory}
I.~P. Fellegi and A.~B. Sunter.
\newblock A theory for record linkage.
\newblock {\em Journal of the American Statistical Association},
  64(328):1183--1210, 1969.

\bibitem{ferreira2012brief}
A.~A. Ferreira, M.~A. Gon{\c{c}}alves, and A.~H. Laender.
\newblock A brief survey of automatic methods for author name disambiguation.
\newblock {\em Acm Sigmod Record}, 41(2):15--26, 2012.

\bibitem{fisher2015clustering}
J.~Fisher, P.~Christen, Q.~Wang, and E.~Rahm.
\newblock A clustering-based framework to control block sizes for entity
  resolution.
\newblock In {\em Proceedings of the 21th ACM SIGKDD International Conference
  on Knowledge Discovery and Data Mining}, pages 279--288. ACM, 2015.

\bibitem{gu2004adaptive}
L.~Gu and R.~Baxter.
\newblock Adaptive filtering for efficient record linkage.
\newblock In {\em Proceedings of the 2004 SIAM International Conference on Data
  Mining}, pages 477--481. SIAM, 2004.

\bibitem{hernandez1995merge}
M.~A. Hern{\'a}ndez and S.~J. Stolfo.
\newblock The merge/purge problem for large databases.
\newblock In {\em ACM Sigmod Record}, volume~24, pages 127--138. ACM, 1995.

\bibitem{hirschman1998muc}
L.~Hirschman and N.~A. Chinchor.
\newblock Muc-7 coreference task definition (version 3.0).
\newblock In {\em Proceedings of MUC-7}, 1998.

\bibitem{kejriwal2013unsupervised}
M.~Kejriwal and D.~P. Miranker.
\newblock An unsupervised algorithm for learning blocking schemes.
\newblock In {\em Proceedings of the IEEE 13th International Conference on Data
  Mining (ICDM)}, pages 340--349. IEEE, 2013.

\bibitem{khabsa2014large}
M.~Khabsa, P.~Treeratpituk, and C.~L. Giles.
\newblock Large scale author name disambiguation in digital libraries.
\newblock In {\em IEEE International Conference on Big Data}, pages 41--42,
  2014.

\bibitem{khabsa2015online}
M.~Khabsa, P.~Treeratpituk, and C.~L. Giles.
\newblock Online person name disambiguation with constraints.
\newblock In {\em Proceedings of the ACM/IEEE Joint Conference on Digital
  Libraries(JCDL'15)}, pages 37--46, 2015.

\bibitem{kim2016inventor}
K.~Kim, M.~Khabsa, and C.~L. Giles.
\newblock Inventor name disambiguation for a patent database using a random
  forest and dbscan.
\newblock In {\em IEEE/ACM Joint Conference on Digital Libraries (JCDL'16)},
  pages 269--270, 2016.

\bibitem{kim2016random}
K.~Kim, M.~Khabsa, and C.~L. Giles.
\newblock Random forest dbscan clustering for uspto inventor name
  disambiguation and conflation.
\newblock In {\em IJCAI-16 Workshop on Scholarly Big Data: AI Perspectives,
  Challenges, and Ideas}, 2016.

\bibitem{levin2012citation}
M.~Levin, S.~Krawczyk, S.~Bethard, and D.~Jurafsky.
\newblock Citation-based bootstrapping for large-scale author disambiguation.
\newblock {\em Journal of the American Society for Information Science and
  Technology}, 63(5):1030--1047, 2012.

\bibitem{liu2014author}
W.~Liu, R.~Islamaj~Do{\u{g}}an, S.~Kim, D.~C. Comeau, W.~Kim, L.~Yeganova,
  Z.~Lu, and W.~J. Wilbur.
\newblock Author name disambiguation for pubmed.
\newblock {\em Journal of the Association for Information Science and
  Technology}, 65(4):765--781, 2014.

\bibitem{mccallum2000efficient}
A.~McCallum, K.~Nigam, and L.~H. Ungar.
\newblock Efficient clustering of high-dimensional data sets with application
  to reference matching.
\newblock In {\em Proceedings of the sixth ACM SIGKDD international conference
  on Knowledge discovery and data mining}, pages 169--178. ACM, 2000.

\bibitem{michelson2006learning}
M.~Michelson and C.~A. Knoblock.
\newblock Learning blocking schemes for record linkage.
\newblock In {\em Proceedings of the 21st AAAI Conference on Artificial
  Intelligence}, pages 440--445. AAAI, 2006.

\bibitem{mooney1995encouraging}
R.~J. Mooney.
\newblock Encouraging experimental results on learning cnf.
\newblock {\em Machine Learning}, 19(1):79--92, 1995.

\bibitem{nadeau2007survey}
D.~Nadeau and S.~Sekine.
\newblock A survey of named entity recognition and classification.
\newblock {\em Lingvisticae Investigationes}, 30(1):3--26, 2007.

\bibitem{papadakis2016comparative}
G.~Papadakis, J.~Svirsky, A.~Gal, and T.~Palpanas.
\newblock Comparative analysis of approximate blocking techniques for entity
  resolution.
\newblock {\em Proceedings of the VLDB Endowment}, 9(9):684--695, 2016.

\bibitem{su2016learning}
G.~Su, D.~Wei, K.~R. Varshney, and D.~M. Malioutov.
\newblock Learning sparse two-level boolean rules.
\newblock In {\em Proceedings of the IEEE 26th International Workshop on
  Machine Learning for Signal Processing (MLSP)}, pages 1--6. IEEE, 2016.

\bibitem{torvik2009author}
V.~I. Torvik and N.~R. Smalheiser.
\newblock Author name disambiguation in medline.
\newblock {\em ACM Transactions on Knowledge Discovery from Data (TKDD)},
  3(3):11, 2009.

\bibitem{yan2007adaptive}
S.~Yan, D.~Lee, M.-Y. Kan, and L.~C. Giles.
\newblock Adaptive sorted neighborhood methods for efficient record linkage.
\newblock In {\em Proceedings of the 7th ACM/IEEE-CS joint conference on
  Digital libraries}, pages 185--194. ACM, 2007.

\end{thebibliography}
	
\end{document}